\documentclass[runningheads]{llncs}
\usepackage{xcolor}
\usepackage[hidelinks]{hyperref}
\hypersetup{
   colorlinks,
   linkcolor={red!50!black},
   citecolor={blue!50!black},
   urlcolor={blue!80!black}
}
\usepackage[LNCS, 
  key=weichselbraun_slot_2022,
  year=2022,
  publication={Natural Language Processing and Information Systems: 27th International Conference on Applications of Natural Language to Information Systems, NLDB 2022, Valencia, Spain, June 15--17, 2022, Proceedings},
  doi={10.1007/978-3-031-08473-7_25},
  doiText={10.1007/978-3-031-08473-7\_25},
  nourl,
  nobib,
  nocopyright
  ]{authorarchive}
\usepackage{booktabs}
\usepackage{algorithm}
\usepackage{algpseudocodex}
\usepackage{graphicx}

\def\makeheadbox{{%
\hbox to0pt{\vbox{\baselineskip=10dd\hrule\hbox
to\hsize{\vrule\kern3pt\vbox{\kern3pt
\hbox{\bfseries [Insert journal name here]}
\hbox{This is a post-peer-review, pre-copyedit version of this article.}
\hbox{The final authenticated version is available online at: \href{[Insert doi here]}{[Insert doi here]}.}
\kern3pt}\hfil\kern3pt\vrule}\hrule}%
\hss}}}

\begin{document}
\title{Slot Filling for Extracting Reskilling and Upskilling Options from the Web}


\author{Albert Weichselbraun \inst{1,2}\orcidID{0000-0001-6399-045X} \and
 Roger Waldvogel \inst{1} \and \orcidID{0000-0001-5191-9816} \and
  Andreas Fraefel \inst{1} \and
 Alexander van Schie \inst{1} \and
 Philipp Kuntschik \inst{1}
}

\authorrunning{A. Weichselbraun et al.}

\institute{University of Applied Sciences of the Grisons, 7000 Chur, Switzerland
 \email{\{firstname.lastname\}@fhgr.ch} \and
 webLyzard technology, 1090 Vienna, Austria
}

\maketitle              

\begin{abstract} 

Disturbances in the job market such as advances in science and technology, crisis and increased competition have triggered a surge in reskilling and upskilling programs. Information on suitable continuing education options is distributed across many sites, rendering the search, comparison and selection of useful programs a cumbersome task.

This paper, therefore, introduces a knowledge extraction system that integrates reskilling and upskilling options into a single knowledge graph. The system collects educational programs from 488 different providers and uses context extraction for identifying and contextualizing relevant content. Afterwards, entity recognition and entity linking methods draw upon a domain ontology to locate relevant entities such as skills, occupations and topics. Finally, slot filling integrates entities based on their context into the corresponding slots of the continuous education knowledge graph.

We also introduce a German gold standard that comprises 169 documents and over 3800 annotations for benchmarking the necessary content extraction, entity linking, entity recognition and slot filling tasks, and provide an overview of the system's performance.
\end{abstract}

\section{Introduction}
\label{sec:introduction}
The automated extraction of structured knowledge from Web content for knowledge base population is a challenging task, since it requires combining content extraction and context aware knowledge extraction.
In the continuing education domain, for example, Web pages promoting courses and degree programs often contain information on learning outcomes and course prerequisites which use entities of similar types (e.g., skills, degrees, etc.). Correctly interpreting these entities requires contextual knowledge of the section in which they appear.

The presented research has been motivated by an industry project, which aims at creating a knowledge graph of national and international educational programs relevant to the Swiss reskilling and upskilling market. To improve the performance of knowledge extraction tasks, the industry partner contributed a comprehensive domain ontology which formalizes domain knowledge such as occupations, skills, topics and positions.

Once completed, the system will cover almost 100,000 educational programs spanning a heterogeneous set of sources such as academic programs, continuing education certificates, courses, seminars, and international online courses. 
The created knowledge graph will power a search platform and recommender system that will support users in locating suitable reskilling and upskilling programs.

The main contributions of this paper are: 
  (i) the application of content extraction and knowledge extraction methods to a complex industry-driven setting that requires building a knowledge graph of educational offerings suitable for reskilling and upskilling;
  (ii) the introduction of methods capable of performing the required complex slot filling task on Web pages retrieved from numerous different education providers;
  (iii) the integration of background knowledge available in the industry partner's domain ontology with state-of-the-art content extraction and knowledge extraction methods;
  (iv) the creation of an infrastructure for assessing the overall task and subtask performance which can be used for benchmarking slot filling on educational content. This infrastructure comprises a German gold standard dataset that contains two partitions of 169 and 75 documents, and six evaluation tasks that consider content extraction, entity extraction, and slot filling.

\section{Related Work}
\label{sec:related}

In recent years, industry-scale knowledge graphs such as the Google Knowledge Graph, Microsoft's Bing knowledge graph, and the knowledge graphs deployed by Facebook, eBay and IBM Watson have emerged \cite{noy_industryscale_2019}. Creating and maintaining such comprehensive knowledge graphs requires significant resources, which has further accelerated research in automated knowledge extraction methods. 

Knowledge base population, for instance, applies knowledge extraction techniques to discover facts in unstructured textual resources to integrate them into a knowledge base or knowledge graph. DBpedia, for example, is constructed by extracting knowledge from Wikipedia Web pages and storing them in the form of (subject, predicate, object) triples \cite{lehmann_dbpedia_2015}. Other approaches operate on more heterogeneous document collections, as reflected in the composition of evaluation datasets covering News articles \cite{lin_canonicalization_2019}, question answering \cite{dubey_lcquad_2019}, and even general Web documents \cite{glass_dataset_2018}. 

Slot filling is a knowledge extraction technique that extracts information on predefined slots (e.g., a person's occupation, age, etc.). When applied to open-world scenarios, slot filling is very challenging, as demonstrated by the TAC 2017 Cold Start Slot Filling Task in which even the winning systems only obtained F-measures below 20\% \cite{lim_unist_2017}. If applied to a single domain, considerably better scores are achieved, as demonstrated by Ritze et al. \cite{ritze_profiling_2016} who use slot filling for augmenting knowledge bases. Most state-of-the-art systems such as Coach \cite{liu_coach_2020}, RSZ\cite{shah_robust_2019} and LEONA \cite{siddique_linguisticallyenriched_2021} deploy deep learning to improve system performance, and transfer learning to allow adaptation to new domains.

The research introduced in this paper, in contrast, customizes slot filling to the target domain by combining deep learning with domain knowledge encoded in a skill and education ontology.


\section{Method}
\label{sec:method}
The presented knowledge extraction system populates the continuing education knowledge graph by drawing upon a skill and education ontology to extract knowledge on educational offerings from websites published by educational institutions such as universities, schools, and course providers.

\subsection{System Overview and Formalization}
\label{sec:system-overview}
\begin{figure}
\centering
\includegraphics[width=.95\linewidth]{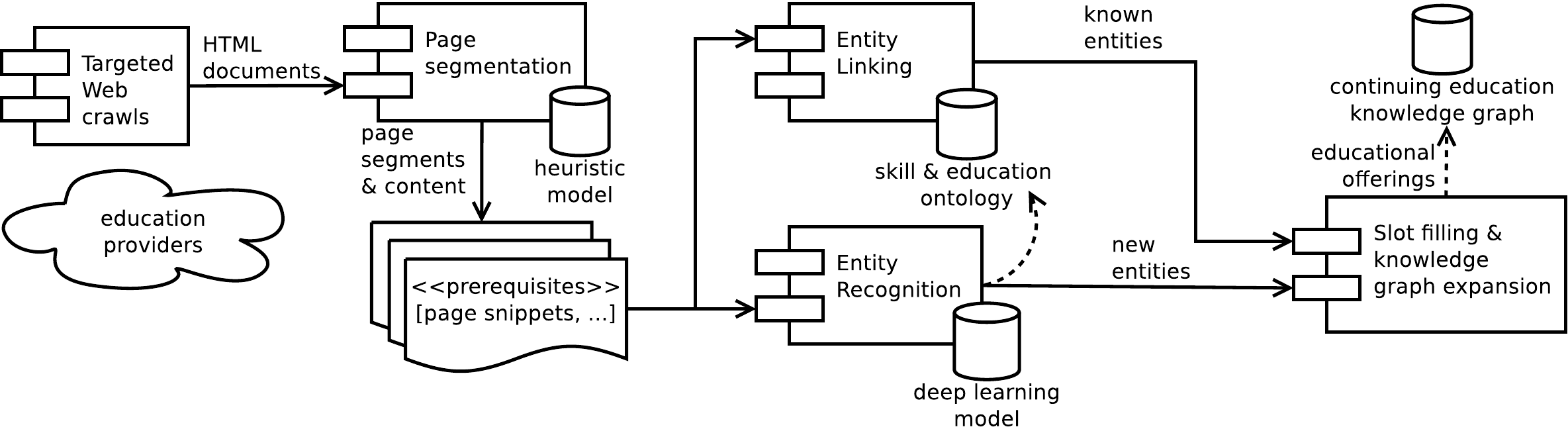}
\caption{\label{fig:system-overview}Overview of the automatic knowledge base population process.}
\end{figure}

Figure~\ref{fig:system-overview} illustrates the developed knowledge extraction and knowledge base population process, which operates on Web pages from 488 different education providers.

The system's knowledge graph population pipeline expands the continuing education knowledge graph by analyzing the Web pages $doc_i$ of educational offerings $i$. A page segmentation and classification component extracts relevant page segments $seg^{type}_{i} \in doc_i$ with $type \in $ \emph{\{target group,} \emph{prerequisite,} \emph{learning objective,} \emph{course content,} \emph{certificates \& degree\}} from these pages. 
Afterwards, entity linking identifies known entities $e^{known}_{ij}$ such as skills and occupation within the segments and links them to the skill and education domain ontology. 
We complement entity linking with entity recognition which is capable of identifying entities $e^{new}_{ij}$ that are not yet available in the domain ontology and, therefore, significantly improves the coverage of the entity extraction process.

\begin{figure}
\centering
\includegraphics[width=.55\linewidth]{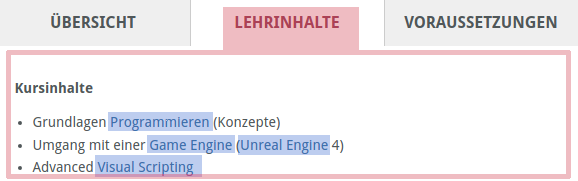}
\caption{\label{fig:example-description}Annotated example course description taken from sae.edu. Blue highlighting indicates identified entities, and the red border outlines the corresponding  page segment.}
\end{figure}

Finally, the {\bf slot filling component} fills for each educational offering $i$ the slots outlined in Table~\ref{tab:slots} by contextualizing the extracted entities $e_{ij} = e^{known}_{ij} \cup e^{new}_{ij}$ with the information on the page segment $seg^{type}_i$ from which they have been extracted. 
New entities that have been discovered by the ER component are assigned unique identifiers that can be used for linking them to the domain ontology at a later stage.

In the real-world example shown in Figure~\ref{fig:example-description}, the slot value with surface form ``Programmierung'' ({\tt edu:prog}) extracted from the page segment ``Lehrinhalte'' ({\tt edu:course\_content}), for example, fills the course content slot, while the same entity extracted from the segment `Voraussetzungen'' ({\tt edu:prerequisites}) would be considered a prerequisite for visiting the course.

\begin{table}[htb]
    \centering
    \caption{\label{tab:slots}Target slot, valid entity types and cardinalities of course entities.}
    \begin{tabular}{llc} \toprule
        slot &  (entity) type & cardinality\\
             & & (min, max) \\ \midrule
        school & school & (1, 1) \\
        target group & topic, occupation, degree, education, industry, position & (0, *) \\
        prerequisite & topic, skill, occupation, position, education & (0, *) \\
        learning objective & topic, skill, occupation & (1, *) \\
        course content & topic, skill & (1, *) \\
        certificates & degree, education & (0, *) \\ \bottomrule
    \end{tabular}
\end{table}

After the completion of the slot filling process, the system integrates the information on each educational offering into the {\bf knowledge graph} by forming the corresponding triples (e.g., {\tt <https://sae.edu\#c01>} {\tt edu:content} {\tt edu:prog.}).

\subsection{Knowledge extraction}
\label{sec:knowledge-extraction}

\subsubsection{Page segmentation.}

 Most websites structure educational offerings in sections, such as course prerequisites and learning objectives. Extracting the content from these sections and correctly labeling it allows contextualization of mined entities, which helps to distinguish slots such as prerequisites, learning outcomes and certificates.
 
 The developed page segmentation system deploys the following three-step process for identifying page segments within the documents:
 (i) \emph{Text segmentation} compares the document's HTML structure with the corresponding text representation generated by the Inscriptis content extraction framework \cite{weichselbraun_inscriptis_2021}. Based on the HTML structure, the text is split into segments whenever separator elements $m = $\emph{\{div, p, li, td, th, dt, dd, summary, legend, h1, h2, h3, h4, h5, h6\}}, which enclose a text and usually do not contain any other element $m$, occurs. 
(ii) \emph{Titles and text clusters} are identified by combining HTML elements with a fallback heuristic which annotates non-standard titles based on their length (max. 3 words) and a task-specific list of commonly used title terms (e.g., `prerequisite', `content', and `degree') obtained from the domain-ontology.
 After identifying the title element, it is used to determine the text cluster by merging all segments until the next title is reached.
 (iii) The final \emph{cluster classification} step compares the title terms with word sequences from the skill \& education ontology to determine the cluster type.

\subsubsection{Entity linking.}
The system uses a graph-based entity linking approach \cite{weichselbraun_mining_2018} that draws upon the project's skill and education database.
We customized the component to differentiate between the project-specific entity types (i) education, (ii) function (i.e., occupation and position), (iii) skill and (iv) topic.

\begin{figure}
\centering
\includegraphics[width=.95\linewidth]{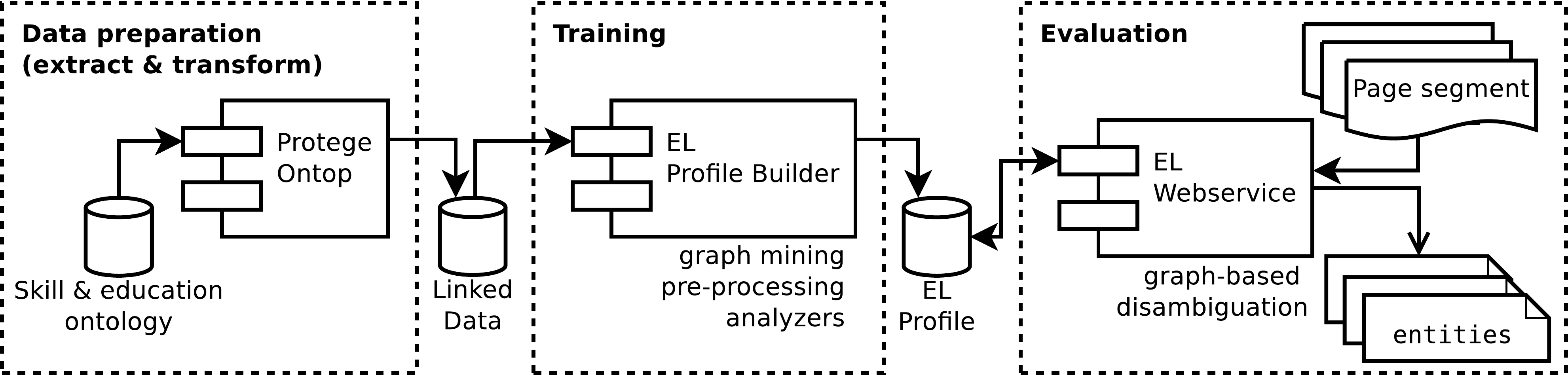}
\caption{\label{fig:el}Overview of the EL process: data preparation transforms the data into a Linked Data repository, training mines the data repository to create a serializable EL profile, and evaluation uses the profile to annotate new and unknown documents.}
\end{figure}

Figure \ref{fig:el} illustrates the utilization of the industry partner's skill and education ontology within the EL component.
The ontology covers most areas relevant to the human resource sector by organizing domain knowledge into 42 tables which combine custom schemas with industry standards such as the \emph{Standard Classification of Occupations}\footnote{https://www.bls.gov/soc/} and multiple industry directories.
The given EL task utilizes tables covering classes, relations and instances on skills, educations, occupations, topics, industries and schools. 
The system's data preparation step deploys Protege Ontop\footnote{https://github.com/ontop/ontop}, a framework that allows translating SPARQL into SQL queries, to transforms these relational data into a Linked Data repository. 

The training step utilizes SPARQL queries to mine, pre-process, and classify relevant entities and context information from the created Linked Data repository. 
The EL Profile Builder further applies multiple pre-processors and analyzers to the query results which create artificial name variations such as plural, possessive forms and abbreviations to maximize the EL profile's coverage, and determine whether the mined surface forms are unambiguous (e.g., ``hedge fund manager''), ambiguous (e.g., ``wolf of wallstreet'') or provide additional context (e.g., ``occupation''). The EL Profile Builder concludes the training by serializing an EL profile containing all information required for the EL process. The EL Web service deploys the profile in conjunction with a graph-based disambiguation algorithm \cite{weichselbraun_mining_2018} that identifies mentions of entities and links them to the ontology concepts.

\subsubsection{Entity recognition and entity classification}
serves as a fallback for entities that have not yet been included into the industry partner's skill and education ontology. 
Entity recognition not only provides an efficient mitigation strategy for identifying these missing entities, but also suggests concepts for inclusion into the skill \& education ontology, which helps in improving its coverage over time.

\begin{figure}
\centering
\includegraphics[width=.95\linewidth]{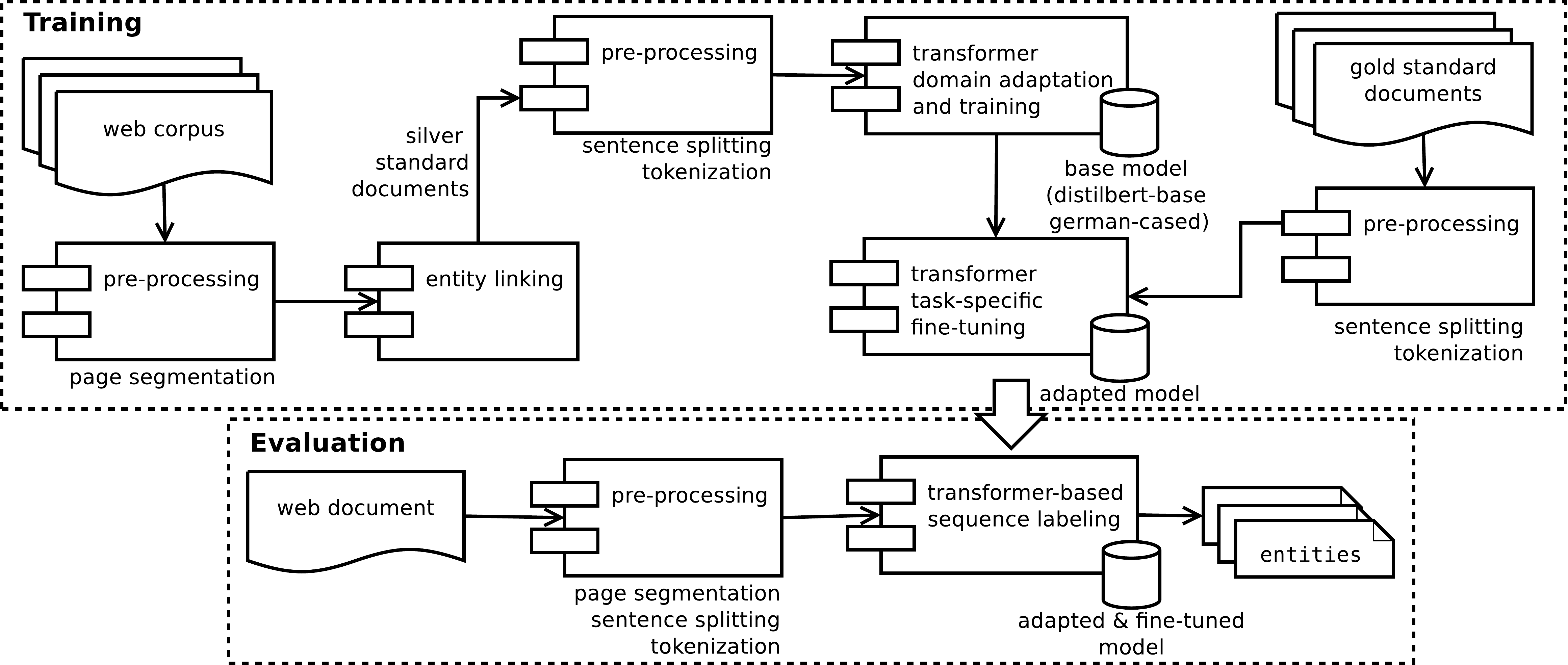}
\caption{\label{fig:transformer}Transformer model used for entity recognition and entity classification.}
\end{figure}

We model entity recognition and entity classification as a token classification problem, where a deep learning model takes a sequence of tokens (e.g., a sentence) and then provides labels for each of them. The entity classification component draws upon the \emph{distilbert-base-german-cased} model provided by the popular transformers library \cite{wolf_transformers_2020}, using the Adam solver with a learning rate of $5 \cdot 10^{-5}$, and a dropout of 0.1. Figure~\ref{fig:transformer} outlines the model structure and the training process.

The domain adaptation step draws upon a domain corpus of 28,000 documents that has been enriched with silver standard annotations obtained from the previously described EL component. Afterwards, fine-tuning draws upon the gold standard documents to further improve the model's capabilities of capturing new entities and to generate the final model. During benchmarking, a five-fold cross evaluation procedure ensures that no training documents are used within the evaluation step.



\section{Evaluation}
\label{sec:evaluation}

\subsection{Gold standard}
\label{sec:gold-standard}
The CareerCoach 2022 gold standard is available for download in the NIF and JSON format\footnote{https://github.com/fhgr/careercoach2022}, and draws upon documents from a corpus of over 99,000 education courses which have been retrieved from 488 different education providers. The project's industry partner classified 20 of these education providers ``high priority'', since they play a key role in the Swiss continuing education market. The list of high priority education providers contains ten academic institutions (universities and universities of applied sciences), five companies focusing on IT education certification, and the five largest education providers not fitting into any of these categories.

\subsubsection{Gold standard partitions.}
\label{sec:partitions}
The corpus contains two partitions. Partition (P1) supports the content extraction tasks and comprises 169 documents and gold standard annotations for page segments. Since the structure of HTML pages frequently differs even within educational offerings, this corpus contains up to five documents per website and also considers a higher total number of education providers.
P1 has been created by sampling (i) five random documents per ``high priority'' education provider and (ii) one random document taken from a selection of 69 randomly selected standard priority education providers.

The second partition (P2) only contains 75 documents but also a significantly richer set of annotations that considers page segments, entities and slots. It, therefore, supports benchmarking knowledge extraction tasks such as entity extraction and slot filling on top of the content extraction task. Since these tasks draw upon the extracted and normalized content, a higher document variability within an education provider seems to be less beneficial, particularly when considering that creating this much richer set of annotations is also very time intensive.

\subsubsection{Annotation guidelines.}
The gold standard annotation process involved two researchers and two industry experts who also outlined guidelines specifying:
(i) \emph{relevant classes} for the page segment recognition and classification tasks, and information on how to identify them; and
(ii) the \emph{entities} supported by the entity extraction tasks and detailed instructions for identifying and distinguishing these entities. This has been particularly important for the 'topic' and 'skill' entity types, which provide valuable context information but are less intuitive to annotators. Entities are mapped to unique identifiers, and the entities of type 'occupation', 'degree', 'education', 'industry', 'school' and 'position'  have also been linked to the corresponding \emph{DBpedia}, \emph{Wikidata} and \emph{European Skill/Competences, qualifications and Occupations (ESCO)}\footnote{https://ec.europa.eu/esco/portal/download} concepts.

During the annotation process, the guidelines have been revisited and further optimized. We drew upon the introduced knowledge extraction pipeline to create candidate annotations that have then been manually corrected and improved by domain experts. In a final step, two additional domain experts from the industry partner manually validated the gold standard annotations based on 28 documents that comprised 1533 annotations, reported potential problems and triggered further improvements to the gold standard and annotation guidelines. 
At this point, the inter-rater-agreement between domain experts from the research institution and the industry experts for partition P2 has been 0.85 (Cohen's Kappa), and 0.88 (pairwise F1 score) \cite{brandsen_creating_2020} for annotated entities. Within the sample, the experts agreed on all page segment annotations, suggesting that the page partitioning task can be easily performed by humans based on the provided annotation guidelines. The substantial inter-rater-agreement also indicates that pre-annotating the corpus is unlikely to introduce any significant bias.

\subsubsection{Gold standard properties.}

Table~\ref{tab:c-properties} provides basic statistics and summarizes information on the number of annotated classes, and annotations per document. The table also gives an impression on the differences to expect between education providers. For instance, there are documents that do not contain any relevant contents (i.e., no relevant page segment) as well as cases where page segments appear multiple times. The entity extraction and slot filling tasks are also confronted with a large variety of entities, ranging from 2 entities per document to a total of 235 entities identified in a single document.

\begin{table}[ht]
\caption{Corpus properties} 
\centering 
\begin{tabular}{lcc}
\hline
     & P1: Content  & P2: Entity extraction \\
     & extraction   & and slot filling  \\ [0.5ex]
\hline
Number of documents & 169 & 75 \\ 
Number of education providers & 89 & 55 \\

Number of annotation classes & 5 & 8 \\
Min. annotations in document & 0 & 2 \\
Avg. annotations in document & 3.4 & 43.9 \\
Median annotations in document & 3 & 30 \\
Max. annotations in document & 17 & 235 \\
\hline 
\end{tabular}
\label{tab:c-properties} 
\end{table}

\subsection{Evaluation tasks}
The evaluation draws upon the gold standard and supports benchmarking the following six tasks performed by the introduced knowledge extraction system:

\subsubsection{Content extraction.}
Content extraction identifies and classifies text segments relevant for the slot filling tasks. We distinguish between,
\begin{enumerate}
    \item \emph{T1: page segment recognition} - identifies page segments within HTML pages $doc$ and extracts the text strings $seg^j_i \in doc_i$ from these segments.
    \item \emph{T2: page segment classification} - assigns each extracted text segment $seg^j_i$ to a class $type \in$ \{\emph{target group}, \emph{prerequisite}, \emph{learning objective}, \emph{course content}, \emph{certificates \& degree\}}.
\end{enumerate}

We evaluate the page segment recognition task (T1) by comparing the tokens in the extracted page segments $t_i \in seg^j_i$ with the tokens in the gold standard segments $t_g \in s^j_g$. 
The evaluation of the page segment classification also requires that the types of the gold standard page segment $t_g^{type}$ and of the extracted segment $t_g^{type}$ match.

\subsubsection{Entity extraction.}
The entity extraction tasks aim at identifying mentions of entities of type $tp_i \in \mathcal{T}_i$ within the extracted page segments. The corpus contains annotations of the following entity types: 'skill', 'occupation', 'topic', 'position', 'school', 'industry', 'education', 'degree'.
\begin{enumerate}
    \item \emph{T3: entity recognition} - locates mentions $m_i$ of entities within text segments.
    \item \emph{T4: entity classification} - assigns each mention $m_i$ to the corresponding entity type $tp_i \in \mathcal{T}_i.$
    \item \emph{T5: entity linking} - links mentions $m_i$ to the appropriate entity $e_i$ in the knowledge graph $KG$. Entities that are not yet available in the knowledge graph are handled as NIL entities (i.e., they are assigned a temporary identifier that is unique for all mentions which refer to the same entity).
\end{enumerate}

We distinguish between two evaluation settings: strict and relaxed. In the strict settings, mentions $m_i$ identified by the entity recognition component for task T3 are considered true positives (TP), if they are identical to a gold standard mention $m_g$.
The entity classification task (T4) also requires that both entities have been assigned to the same entity type $tp_i$, and the linking task (T5) requires linking the mention to the correct knowledge base entity $e_i$.

The \emph{relaxed setting} eases these conditions by also considering mentions that overlap a gold standard mention as correct.

Entities that do not appear in the gold standard are considered false positives (FP), and false negatives (FN) refer to gold standard entities which have been missed by the entity extraction task.

\subsubsection{Slot filling.}
The slot filling task (T6) combines all the tasks above. The page segment recognition (T1) identifies page segments. Afterwards, the page segment classification (T2) assigns them to the corresponding segment cluster, and entity recognition (T3), entity classification (T4) and entity linking (T5) are performed. 
Finally, we assign the extracted entities $e_i$, which have been contextualized based on the classification of the page segment in which they have been identified, to the corresponding slot.


\subsection{Experiments and discussion.}
Table~\ref{tab:results} summarizes the results of a five-fold cross evaluation on the CareerCoach 2022 gold standard. For the entity linking and slot filling task, the evaluation also distinguishes between the strict and the relaxed setting. 

\begin{table}
  \centering
  \caption{Slot filling and per component evaluation results.}
  \label{tab:results}
  \begin{tabular}{ll@{\hskip 0.5cm}l@{\hskip 0.5cm}l}
  \toprule
  Component & P & R & F1  \\
  \midrule
  T1: page segment recognition & 0.82 & 0.84 & 0.83\\
  T2: page segment classification & 0.82 & 0.84 & 0.83\\
  T3: entity recognition & 0.82 & 0.66 & 0.73 \\
  T4: entity classification & 0.78 & 0.63 & 0.70  \\
  T5: entity linking (strict) & 0.67 & 0.80 & 0.73  \\
  T5: entity linking (relaxed) & 0.67 & 0.82 & 0.74 \\
  T6: slot filling (strict) & 0.48 & 0.60 & 0.54  \\
  T6: slot filling (relaxed) & 0.50 & 0.62 & 0.55 \\\bottomrule
  \end{tabular}
\end{table}


A comparison of the page segment recognition and page segment classification performance reveals the same scores for both tasks. This confirms that the developed simple segment classification heuristic has been very effective and has classified all page segments correctly.

The results indicate that both entity recognition and entity classification have been optimized towards a higher precision, to spare domain experts, which need to confirm new entity types in the production process. 
Entity linking, in contrast, has been optimized towards recall, as outlined in the evaluation results.

The evaluation of the overall slot filling process only considers correctly assigned slots (i.e., course, slot and slot value are correct) as true positives. All other values are considered false positives, and missing values as false negatives. The F1 score of the slot filling process indicates that the system is not yet suitable for fully automated knowledge population, but rather enables a semi-automated process that significantly improves throughput when compared to the prior deployed manual approaches. 

\subsection{Automatic Knowledge Graph Population}
Running the presented system on 55 course descriptions (one per unique education provider) from the gold standard, extends the knowledge base by 453 unique statements. 
Most of these statements (222) describe the course content, followed by target groups (90), learning objectives (61), course prerequisites (51) and certificates (29). In addition, 511 slot values have been marked as ``related'' since the system hasn't been able to unequivocally resolve their slot, due to shortcomings in the page partitioning process. This result indicates, that further improving the page partitioning process will be key towards enhancing the system's recall.

\begin{figure}
\begin{verbatim}
<https://sae.edu#c01>  edu:prerequisite    edu:matura, edu:student;
                       edu:course_content  edu:Design, edu:Podcasting;
                       skos:related        edu:Web_Design, edu:Publishing.\end{verbatim}
\caption{\label{fig:kg}Snippet of six RDF triples that have been integrated into the knowledge graph.}
\end{figure}

\section{Outlook and conclusions}
\label{sec:outlook}
Performing slot filling tasks on third-party websites is a challenging task, requiring content extraction and knowledge extraction methods to work in concert.

This paper introduces a slot filling system that mines education provider websites for a wide range of educational offerings such as academic programs, continuing education programs, courses, seminars and online courses. Integrating background knowledge from a proprietary ontology allows the application of graph-based entity linking methods for the identification of known entities,  which are complemented by entity recognition to mitigate coverage issues within the ontology. The slot filling component contextualizes entities to distinguish between ambiguous slots such as \emph{prerequisites} versus \emph{learning outcomes}, and afterwards integrates them into a continuing education knowledge graph.

An evaluation framework comprising six evaluation tasks and a publicly available German gold standard allow benchmarking the content extraction and knowledge extraction methods utilized within the slot filling system. 
The framework did not only yield information on the components' performance, but also guided the system development by providing rapid feedback on the impact of changes and improvements. In addition, it offers a reliable benchmark to third-party researchers interested in the described slot filling task.

Future work will focus on: (i) improving the slot filling performance by enhancing page segmentation, increasing the coverage of the proprietary knowledge graph used for entity linking, and fine-tuning the entity recognition component. Given the importance of the created benchmarking framework for the research and development process, we plan on (ii) further increasing its size and coverage; and (iii) integrating the gold standard with explainable benchmarking frameworks such as Orbis \cite{odoni_introducing_2019} to make it more accessible to third-party researchers.


\subsection*{Acknowledgments}
The research presented in this paper has been conducted within the CareerCoach project (www.fhgr.ch/CareerCoach) which is funded by Innosuisse.


\end{document}